\documentclass[twocolumn,floatfix, showpacs]{revtex4}

\usepackage[final]{epsfig}
\usepackage{amsmath}
\usepackage{parskip}
 
\begin{document}
 
\title{$\gamma$-hadron correlations as a tool to trace the flow of energy lost from hard partons in heavy-ion collisions}
 
\author{Thorsten Renk}
\email{trenk@phys.jyu.fi}
\affiliation{Department of Physics, P.O. Box 35 FI-40014 University of Jyv\"askyl\"a, Finland}
\affiliation{Helsinki Institute of Physics, P.O. Box 64 FI-00014, University of Helsinki, Finland}
 
\pacs{25.75.-q,25.75.Gz}
\preprint{HIP-2006-46/TH}

\begin{abstract}
High transverse momentum ($P_T$) $\gamma$-hadron correlations are currently being regarded as the 'golden channel' for the study of the medium produced in ultrarelativistic heavy ion collisions by means of hard probes. This is due to several reasons, all linked to the fact that because of the smallness of the electromagnetic coupling $\alpha$, the photon does not substantially interact with the medium and is expected to escape unmodified. Thus, a high $P_T$ photon indicates a hard process in the collision independent of the position of the hard vertex. In contrast, there may not be a clear signal for a hard process involving strongly interacting partons if the production vertex is deep in the medium as both partons undergo substantial final state interaction. Equally important, if photon production by fragmentation can be separated experimentally, the photon provides almost full knowledge of the initial kinematics. In the present paper, these properties are used to demonstrate a distinguishing feature between two assumptions made in modelling the medium-modifications of strongly interacting high $P_T$ processes: Loss of energy into the medium vs. medium modification of the partonic shower. Is it shown that $\gamma$-h correlations provide a very clean signature to distinguish the two scenarios. 
\end{abstract}
 
\maketitle

\section{Introduction}

The main goal of studying medium-induced modifications to hard processes in heavy-ion (A-A) collisions is to do jet tomography on the medium \cite{Jet1,Jet2,Jet3,Jet4,Jet5,Jet6}. The underlying idea is as follows: Since hard processes take place slightly before the creation of a soft bulk medium, high $p_T$ partons have to traverse the medium before hadronization, and interactions with the soft medium subsequently lead to an energy redistribution, resulting in a suppression of observed hard hadron yield. The strength of the interaction with the medium reflects the density of the medium, thus one should be able to gain information about the medium density from the observed strength of the suppression. The chief observable considered so far has been the nuclear suppression factor of single hadrons $R_{AA}$ (cf. e.g. \cite{PHENIX,STAR-data}) which is the measured yield in A-A collisions divided by the yield in proton-proton (p-p) collisions multiplied with the number of binary scatterings (i.e. the default expectation if there would not be soft processes forming a medium in an A-A collision).

There are at the moment two main scenarios which are considered in models to explain the measured suppression: Energy loss and the medium-modified parton shower. In the energy loss picture, the approximation is made that a single parton carries most of the energy and momentum of the event. Note that this approximation is justified for single inclusive hadron spectra or back-to-back correlations, but clearly not for fully reconstructed jets. In traversing the medium, this leading parton loses energy to the medium and subsequently fragments into a hadron shower in vacuum after emerging from the medium with reduced energy. This picture describes the available data well \cite{HydroJet1,Dihadron1,Dihadron2} using e.g. the Armesto-Salgado-Wiedemann (ASW) implementation \cite{QuenchingWeights} of energy loss. The energy loss model itself does not explicitly specify what happens to the lost energy (it is computed in the form of gluon radiation, but clearly also radiated gluons would re-interact with the medium), but it is often assumed that lost energy excites shockwaves in a hydrodynamical bulk medium, see e.g. \cite{MachShuryak,Mach1,Mach2,Mach3,Source}. 

In contrast, the medium-modified parton shower picture is not restricted to the approximation that a single parton carries most of the energy and momentum, as it follows medium-modified evolution equations for the full sequence of branchings of the initially produced high virtuality parton. Only the nonperturbative part of fragmentation, i.e. hadronization, is assumed to take place outside the medium. There is a class of models in which this evolution of a whole parton distribution is followed analytically \cite{HBP,HydroJet2,HydroJet3,Dihadron3} and also Monte-Carlo (MC) descriptions of the in-medium shower \cite{JEWEL,YaJEM1,YaJEM2,YaJEM3,Carlos} based on the vacuum shower evolution routines in PYTHIA \cite{PYTHIA} or HERWIG \cite{HERWIG}. While these models can account for $R_{AA}$ well, usually the detailed comparison with more differential observables is not quite as advanced yet due to the greater complexity of these models. The important difference to the energy loss picture is that in the medium-modified parton shower picture, the flow of energy from the leading parton to subleading partons is explicitly traced. Thus, such models make a very clear prediction that the energy lost from the leading shower partons leads to increased parton (and subsequently hadron) production at lower momenta in the shower.

$\gamma$-hadron correlations \cite{XNPhotons1,XNPhotons2,MyPhotons,XNPhotons3} offer a way to distinguish the two scenarios.  If the medium-modified parton shower picture is the correct way to describe the energy redistribution due to the medium, the additional multiparticle production at low $P_T$ can be experimentally found as a substantial enhancement of the per trigger yield scaled to the per trigger yield in p-p collisions ($I_{AA})$ in a comparatively narrow angular region (say $\alpha < \pi/5$) back to the photon at low $z_T$ (where for each hadron $h$ the relation $E_{h} = z_T E_\gamma$ links hadron energy $E_h$ and photon energy $E_\gamma$). Here, the photon serves both as an indicator that a hard process has taken place and to identify the kinematics of that process. On the other hand, if the energy is redistributed as a shockwave in the medium \cite{MachShuryak,Mach1,Mach2,Mach3,Source}, any correlation of hadrons with the photon will be at thermal momenta and at large angles, thus it will {\em not} lead to an enhancement of $I_{AA}$ when only an angular region of $\alpha < \pi/5$ is considered to extract the per-trigger yield of hadrons correlated with the photon. However, in both scenarios an $I_{AA}$ below unity is expected at high $z_T$ since this is the kinematical region from which energy is lost and which is also probed in $R_{AA}$.

In this paper, we present a systematic comparison of $I_{AA}$ computed both in the energy loss picture and the medium-modified parton shower picture. For the energy loss picture, we use the ASW quenching weights \cite{QuenchingWeights}. Note that this formalism is very well constrained by other data \cite{HydroJet1,Dihadron1,Dihadron2}. $I_{AA}$ in the medium-modified shower picture is computed using the MC code YaJEM (Yet another Jet Energy-loss Model) \cite{YaJEM1,YaJEM2,YaJEM3} using the ratiative energy loss (RAD) setting. Both calculations are done using the same 3-d hydrodynamical model of the bulk medium evolution \cite{Hydro3d} which is constrained by a large number of bulk observables. In the following, we consider only direct photons and assume that photons from fragmentation processes which would otherwise be a small modification to the results can be separated experimentally, either statistically or by means of isolation cuts.

\section{The model}

The model has four main building blocks: 1) the primary hard photon and parton production, 2) the propagation of the away side partons through the medium and the parton-medium interaction 3) the evolution of the bulk medium and 4) the hadronization of partons emerging from the medium. Steps 1) and 3) are common to both the energy loss and the medium-modified parton shower picture, but 2) and 4) will be described separately for the energy loss picture in the ASW formalism and for the medium-modified parton shower picture in YaJEM. 3) is computed in the 3-d hydrodynamical model by Bass and Nonaka which is described elsewhere \cite{Hydro3d}. In the hydrodynamical model, thermodynamical properties like the medium energy density $\epsilon$ or its temperature $T$ can be computed at all spacetime points.

The framework in which all the in ingredients are merged is a MC code for the generation of back-to-back events in the medium which has been used extensively for the computation of hard dihadron correlations \cite{Dihadron1,Dihadron2,Dihadron4,Dihadron5} and is described there in detail. The presentation here will just outline the essential physics issues.

\subsection{Primary parton production}

The production of a hard partons $k$ and a photon $\gamma$ in leading order (LO) perturbative Quantum Choromdynamics (pQCD) is described by
 
\begin{equation}
\label{E-2Parton}
\frac{d\sigma^{AB\rightarrow k\gamma +X}}{d p_T^2 dy_1 dy_2} \negthickspace = \sum_{ij} x_1 f_{i/A} 
(x_1, Q^2) x_2 f_{j/B} (x_2,Q^2) \frac{d\hat{\sigma}^{ij\rightarrow k\gamma}}{d\hat{t}}
\end{equation}
 
where $A$ and $B$ stand for the colliding objects (protons or nuclei) and $y_{1(2)}$ is the 
rapidity of parton $k$ and the photon $\gamma$. The distribution function of a parton type $i$ in $A$ at a momentum 
fraction $x_1$ and a factorization scale $Q \sim p_T$ is $f_{i/A}(x_1, Q^2)$. The distribution 
functions are different for the free protons \cite{CTEQ1,CTEQ2} and nucleons in nuclei 
\cite{NPDF,EKS98}. The fractional momenta of the colliding partons $i$, $j$ are given by
$ x_{1,2} = \frac{p_T}{\sqrt{s}} \left(\exp[\pm y_1] + \exp[\pm y_2] \right)$.

By selecting pairs $k,\gamma$ and summing over all possible combinations of initial partons $i,j$ for the two contributing channels $q\overline{q} \rightarrow \gamma g$ and $q g \rightarrow q \gamma$ and  where $q$ stands for any of the quark flavours $u,d,s$
we find the relative strength of the production channels as a function of $p_T$. In the kinematic range probed at RHIC, $qg \rightarrow q\gamma$ dominates, thus quark jets are predominantly correlated with a photon trigger.

For the present investigation, we require $y_1 = y_2 = 0$, i.e. we consider only back-to-back correlations detected at midrapidity. In a first step, we sample Eq.~(\ref{E-2Parton}) summed over all $k,\gamma$ to generate $p_T$ for the event, in the second step we perform a MC sampling of the decomposition of Eq.~(\ref{E-2Parton}) according to the photon production channel to determine if the parton recoiling from the photon is a quark or a gluon.

To account for various effects, including higher order pQCD radiation, transverse motion of partons in the nucleon (nuclear) wave function and effectively also the fact that hadronization is not a collinear process, we fold into the distribution an intrinsic transverse momentum $k_T$ with a Gaussian distribution, thus creating a momentum imbalance between the photon and the parton as ${\bf {p_T}_\gamma} + {\bf {p_T}_k} = {\bf k_T}$.

Assuming that the photon does not interact further with the medium, we test at this point if the photon momentum falls into a specified trigger range. If not, a new event is generated, otherwise the event is processed further by computing the fate of the away side parton. For this, first the production point in the transverse plane must be found.

The probability density $P(x_0, y_0)$ for finding a hard vertex at the 
transverse position ${\bf r_0} = (x_0,y_0)$ and impact 
parameter ${\bf b}$ is in leading order given by the product of the nuclear profile functions as
\begin{equation}
\label{E-Profile}
P(x_0,y_0) = \frac{T_{A}({\bf r_0 + b/2}) T_A(\bf r_0 - b/2)}{T_{AA}({\bf b})},
\end{equation}
where the thickness function is given in terms of Woods-Saxon the nuclear density
$\rho_{A}({\bf r},z)$ as $T_{A}({\bf r})=\int dz \rho_{A}({\bf r},z)$.  Note that Eq.~(\ref{E-Profile}) may receive (presumably) small corrections when going beyond a leading order calculation.

After the first step, we thus have generated a back-to-back event of a photon and a parton with known kinematics $p_\gamma, p_k$ and initial vertex position in the transverse plane ${\bf r_0} =(x_0,y_0)$. Partons can now be propagated from this position through the medium for a fixed angle with the reaction plane or averaged over all angles. In the present paper, we use the latter option.

\subsection{Parton-medium interaction and hadronization using ASW quenching weights}

For any angle $\phi$ of the parton with the reaction plane, 
the path of the parton through the medium $\zeta(\tau)$ is specified 
by $({\bf r_0}, \phi)$ and we can compute the energy loss 
probability $P(\Delta E)_{path}$ for this path. We do this by 
evaluating the line integrals
\begin{equation}
\label{E-omega}
\omega_c({\bf r_0}, \phi) = \int_0^\infty \negthickspace d \zeta \zeta \hat{q}(\zeta) \quad  \text{and} \quad \langle\hat{q}L\rangle ({\bf r_0}, \phi) = \int_0^\infty \negthickspace d \zeta \hat{q}(\zeta)
\end{equation}
along the path where we assume the relation
\begin{equation}
\label{E-qhat}
\hat{q}(\zeta) = K \cdot 2 \cdot \epsilon^{3/4}(\zeta) (\cosh \rho - \sinh \rho \cos\beta)
\end{equation}
between the local transport coefficient $\hat{q}(\zeta)$ (specifying 
the quenching power of the medium), the energy density $\epsilon$ and the local flow rapidity $\rho$ with angle $\beta$ between flow and parton trajectory. Energy density $\epsilon$ and local flow rapidity $\rho$ are the input from the 3-d hydrodynamical simulation of the medium evolution \cite{Hydro3d}.

$\omega_c$ is the characteristic gluon frequency, setting the scale of the energy loss probability distribution, and $\langle \hat{q} L\rangle$ is a measure of the path-length weighted by the local quenching power.
We view  the parameter $K$ as a tool to account for the uncertainty in the selection of $\alpha_s$ and possible non-perturbative effects increasing the quenching power of the medium (see discussion in \cite{Dihadron2}) and adjust it such that pionic $R_{AA}$ for central Au-Au collisions is described. This leads to a value of $K=3.6$ \cite{HydroJet1}.

Using the numerical results of \cite{QuenchingWeights}, we obtain $P(\Delta E; \omega_c, R)_{path}$ 
for $\omega_c$ and $R=2\omega_c^2/\langle\hat{q}L\rangle$ for given parton production vertex and angle $\phi$. We sample this distribution to determine the actual energy loss of the away side parton in the event and subtract it from the away side parton. The away side parton $k$ is considered completely absorbed by the medium whenever $\Delta E > 0.9 E_k$. If the parton emerges with a finite amount of energy left, it is subsequently hadronized.

Hadronization is modelled by an expansion of the fragmentation function in terms of a tower of conditional probability densities $A_N(z_1, \dots, z_n, \mu)$ with the probability to produce $n$ hadrons with momentum fractions $z_1, \dots z_n$ from a parton with momentum $p_T$  being $\Pi_{i=1}^n A_i(z_1,\dots z_i,p_T)$. The procedure is described in detail in \cite{Dihadron-LHC} where we used PYTHIA \cite{PYTHIA} to simulate the shower, in the present paper we employ HERWIG \cite{HERWIG} instead. We use the exact expressions for the first two terms and the approximations 

\begin{equation}
A_3(z_1, z_2, z_3, p_T) \approx A_2(z_1+z_2, z_3, p_T) \theta(z_2-z_3)
\end{equation}

and

\begin{equation}
\begin{split}
A_4(z_1, z_2, z_3, z_4, p_T) \approx& A_2(z_1+z_2+z_3, z_4, p_T) \\
& \times \theta(z_2-z_3) \theta(z_3 - z_4).
\end{split}
\end{equation}

for the next two terms of the expansion. This expansion only becomes invalid at very low $z$, we will thus exclude this region from the discussion of results.

Thus, for each event in the back-to-back MC code, we have after hadronization up to four hadrons from which we calculate the conditional per-trigger yield given the initial photon. We compute this conditional yield as a function of $z_T$ which is the fraction of the photon momentum taken by each hadron. Note that due to the intrinsic $k_T$ smearing, $z_T$ can be above unity, unlike $z$.

\subsection{Parton-medium interaction using YaJEM}

YaJEM \cite{YaJEM1,YaJEM3} is a tool capable of computing the complete medium modified fragmentation function given a particular path through the medium. For the radiative energy loss scenario (RAD), the medium needs to be characterized by a parameter $\hat{q}$ which is linked to the thermodynamical properties of the medium by means of Eq.~(\ref{E-qhat}), albeit with a different $K=3$ which again results from a best fit to pionic $R_{AA}$ in central Au-Au collisions. 

The main physics assumption of the RAD scenario YaJEM is that the evolution equations for parton branching can be solved with each parton acquiring additional virtuality while propagating through the medium. The gain of parton $a$ in virtuality $\Delta Q_a^2$ is then given by

\begin{equation}
\label{E-Qgain}
\Delta Q_a^2 = \int_{\tau_a^0}^{\tau_a^0 + \tau_a} d\zeta \hat{q}(\zeta).
\end{equation}

This requires that the parton evolution which takes place in momentum space is linked to the medium evolution in position space. In YaJEM, this link is done based on formation time arguments. The average lifetime of a virtual state $b$ coming from a parent state $a$ is computed as 

\begin{equation}
\label{E-Lifetime}
\langle \tau_b \rangle = \frac{E_b}{Q_b^2} - \frac{E_b}{Q_a^2}.
\end{equation}  

and the actual formation time in a given event is be obtained from the probability distribution 

\begin{equation}
\label{E-RLifetime}
P(\tau_b) = \exp\left[- \frac{\tau_b}{\langle \tau_b \rangle}  \right].
\end{equation}

Based on this time information, the spatial position of partons in the medium is obtained by propagating them on an eikonal trajectory given by the shower-initiating parton (thus neglecting the small transverse spread of partons in the shower). The medium-induced virtuality leads to additional branching and hence an energy transfer from high $p_T$ partons into the production of multiple partons at lower $p_T$. For hadronization, the complete parton shower is processed using the Lund model \cite{Lund} which is part of PYTHIA \cite{PYTHIA}.

Due to an approximate scaling law identified in \cite{YaJEM1}, it is sufficient to compute the line integral

\begin{equation}
\Delta Q^2_{tot} = \int d \zeta \hat{q}(\zeta)
\end{equation}

in the medium to obtain the full medium-modified fragmentation function (MMFF) $D_{MM}(z, \mu^2,\zeta)$ for a given eikonal path of the shower-initiating parton. Thus, for each event in the back-to-back MC code, YaJEM can provide the full MMFF, thus this is the quantity we average after transforming it into a function of $z_T$ given the photon energy in the event.

\section{Results}

\begin{figure}
\epsfig{file=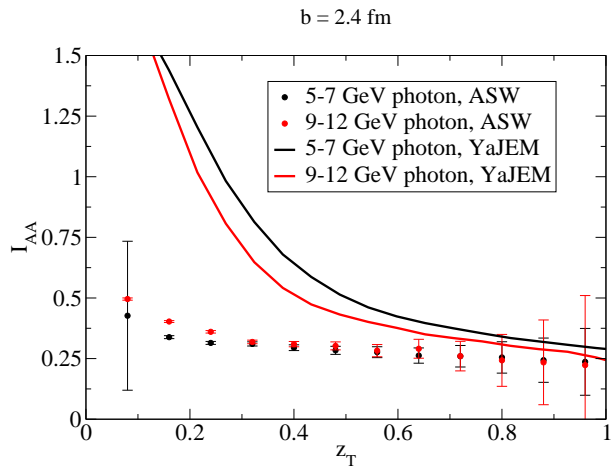, width=8cm}
\caption{\label{F-Comp}Ratio $I_{AA}$ of the per trigger away side yield in central 200 AGeV Au-Au collisions  divided by the result in p-p collisions given a photon trigger in the indicated range as a function of $z_T$. The away side yield is extracted in a region in a cone with $\alpha < \pi/5$ around the direction opposite to the photon. Shown are results in the energy loss picture (ASW) and in the medium-modified parton shower picture (YaJEM).}
\end{figure}

With the two different frameworks for parton-medium interaction and hadronization, we are in a position to compute the conditional yield on the away side given a photon trigger both for p-p and central Au-Au collisions. To exclude the region of very low $P_T$, we only consider hadrons above 0.5 GeV. The result is shown in Fig.~\ref{F-Comp}. Note that there is a technical point to consider: In the ASW calculation, hadronization is computed inside the back-to-back MC simulation and thus both the sampling of geometry and energy loss and the sampling of the fragmentation function is done with the same number of events. In the YaJEM calculation, hadronization is not computed within the back-to-back MC code but within YaJEM and the full MMFF is passed into the back-to-back simulation. Thus, effectively the YaJEM calculation has a factor $\sim$100.000 more statistics and thus vanishing statistical errors whereas the ASW calculation is limited in statistics at large $z_T$ --- that region, however, is not so relevant for the present paper.

It is immediately clear from the figure that there is a dramatic difference between the expectations for $I_{AA}$ for the two different scenarios. While the ASW scenario shows only a moderate rise towards low $z_T$, the YaJEM scenario exhibits a dramatic increase even above unity. However, both scenarios agree in the high $z_T$ region as expected --- the depletion here is in essence the same physics that causes single hadron $R_{AA} < 1$. While there is some small dependence on the photon momentum, to first order the dramatic differences between the scenarios are independent of the trigger range.

Let us first discuss the main feature, the different low $z_T$ behaviour. In any medium-modified shower picture, there is a sum rule --- if there is to be a depletion of high $P_T$ hadrons in an in-medium shower as compared to a shower in vacuum, then energy conservation requires that this is compensated by an increase of low $P_T$ hadrons. Thus, if the fragmentation function ratio is below unity in the high $z_T$ range, it must rise above unity at low $z_T$. This is not so in the energy loss picture where energy is assumed to be distributed to extremely low momenta and large angles --- here the ratio may stay below unity for any value of $z_T$ if the yield is extracted with a momentum cut and the angular restriction on the away side as outlined above. 

Next, let us turn to the small rise of $I_{AA}$ at low $z_T$ in the ASW scenario. This is in essence a kinematical effect. For high $z_T$, one reqires a hadron which has approximately (modulo the intrinsic $k_T$) the full energy of the original parton. Any small amount of energy loss will thus immediately lead to a dramatic suppression, and thus one is in essence observing here hadrons from partons which have not lost any energy. This is not so at low $z_T$ --- since the hadron energy is far from the primary parton energy in this region, there is some room for energy loss, and thus one observes events in which there was a moderate energy loss in addition to events in which the parton did not lose energy, and this is reflected in a rise. Note that the rise is more pronounced for the higher trigger energy where there is even more room for energy loss. In any scenario in which energy loss and in-medium pathlength are strongly linked, as e.g. \cite{XNPhotons3}, this kinematical argument can be translated into a geometrical argument leading to a transition from surface emission (no energy loss) to volume emission (moderate energy loss). However, if energy loss is a strongly fluctuating function of in-medium pathlength (as in the ASW scheme used here) the geometrical interpretation of the kinematical argument breaks down.

Thus, there is a very clear distinction between the two physics pictures combined with a good insight into what causes the difference.

\section{Conclusions}

We have computed $I_{AA}$ the ratio of the per-trigger away side yield in central Au-Au collisions over p-p collisions given a photon trigger as a function $z_T$. The away side yield is extracted in an angular region with $\alpha < \pi/5$ around the direction back-to-back with the photon. $I_{AA}(z_T)$ shows a clear difference between a calculation under the assumption that energy is lost from a hard parton and redistributed in the medium or under the assumption that energy is transferred from hard partons to softer partons inside the shower during the in-medium shower evolution. In the case of energy loss, $I_{AA}(z_T) <1$ is true for all $z_T$ whereas for the medium-modified parton shower $I_{AA}(z_T)>1$ for small $z_T$ is unavoidable.

Preliminary data \cite{gamma-h-talks} from both PHENIX and STAR for $I_{AA}$ in $\gamma$-hadron correlations show no indication of a raise of $I_{AA} > 1$ for any $z_T$ or any trigger momentum. If these results are confirmed, they would constitute strong evidence that the nature of the parton-medium interaction is such that energy lost from an initially hard parton is almost completely absorbed and redistributed in the medium and that pictures based on an in-medium evolution of the shower (which are in principle more complete than energy loss pictures) miss an essential physics process.

\begin{acknowledgments}
 
I would like to thank B.~Jacak, M.~Connors, S.~Mioduszewski and A.~Hamed for interesting discussions and their time to explain to me some of the details of $\gamma$-h correlation measurements.  This work was financially supported by the Academy of Finland, Project 115262.
 
\end{acknowledgments}

\end{document}